\pdfoutput=1
\documentclass{JINST}
\let\ifpdf\relax

\usepackage{commath}

\DeclareMathOperator{\Var}{Var}

\title{Chromatic CCD effects on weak lensing measurements for LSST}

\author{J.~E.~Meyers, P.~R.~Burchat\\
Kavli Institute for Particle Astrophysics and Cosmology, Department of Physics, Stanford University, Stanford CA 94305\\
E-mail: \email{jmeyers3@stanford.edu}}

\abstract{
Wavelength-dependent point spread functions (PSFs) violate an implicit assumption in current galaxy shape measurement algorithms that deconvolve the PSF measured from stars (which have stellar spectral energy distributions (SEDs)) from images of galaxies (which have galactic SEDs).
Since the absorption length of silicon depends on wavelength, CCDs are a potential source of PSF chromaticity.
Here we develop two toy models to estimate the sensitivity of the cosmic shear survey from the Large Synoptic Survey Telescope to chromatic effects in CCDs.
We then compare these toy models to simulated estimates of PSF chromaticity derived from the LSST photon simulator \textsc{PhoSim}.
We find that even though sensor contributions to PSF chromaticity are subdominant to atmospheric contributions, they can still significantly bias cosmic shear results if left uncorrected, particularly in the redder filter bands and for objects that are off-axis in the field of view.
}

\begin{document}

\section{Introduction}\label{sec:intro}

All galaxy shape measurement algorithms for ground-based telescopes depend, in one way or another, on measuring the point spread function (PSF) from the images of stars to then deconvolve from the images of galaxies.
The implicit assumption in this approach is that the PSF affecting stars and galaxies is locally the same.
If the PSF depends on wavelength, however, then this assumption is violated since stars and galaxies have different spectral energy distributions (SEDs).

Several papers have addressed aspects of this effect to date.  
Cypriano et al. (2010) \cite{Cypriano++10} analyze the impact of the wavelength-dependent diffraction limit of the {\it Euclid} PSF, Plazas \& Bernstein (2012) \cite{Plazas+Bernstein12} investigate differential chromatic refraction for ground-based telescope PSFs such as DES and LSST, and Meyers \& Burchat (2015) \cite{Meyers+Burchat15} study both differential chromatic refraction and chromatic seeing for ground-based telescopes.

Chromatic effects are also likely to arise in CCDs due to the wavelength dependence of the absorption length in silicon \cite{O'Connor++06}.
LSST will primarily measure galaxy shapes in the $r$- and $i$-band filters, for which the typical silicon absorption lengths are $\sim$ 2-5 microns and $\sim$ 5-20 microns, respectively, which can be compared to the LSST CCD thickness of 100 microns and pixel width of 10 microns.
In this article, we investigate some consequences of the wavelength-dependent absorption depth on the PSF.

In Section \ref{sec:analytics} we briefly review some analytic formulae for estimating shape measurement biases due to PSF misestimates, including those arising from chromatic PSFs.
We then use these analytic expressions to investigate two toy models of chromatic CCD PSFs in Section \ref{sec:toys}, and roughly estimate the tolerance of LSST weak lensing science to generic chromatic CCD effects.
In Section \ref{sec:phosim} we use the LSST photon simulator package \textsc{PhoSim} to estimate the magnitude of chromatic PSF effects for LSST and present our conclusions in Section \ref{sec:discussion}.

\section{Analytics}\label{sec:analytics}

Weak gravitational lensing is frequently analyzed through its effect on combinations of the second central moments $I_{\mu\nu}$ of a galaxy's surface brightness distribution.
For a detailed description of this process, see Ref. \cite{Meyers+Burchat15}.
We briefly summarize the salient points for the present study here.

The zeroth, first, and second unweighted central moments of the surface brightness profile $I(x,y)$ are
\begin{equation}
  \label{eqn:M0}
f = \int{ I(x,y)\dif{x} \dif{y}},
\end{equation}
\begin{equation}
  \label{eqn:M1}
\bar{\mu} = \frac{1}{f}\int{ I(x,y)\mu\dif{x} \dif{y}},
\end{equation}
and 
\begin{equation}
  \label{eqn:M2}
  I_{\mu \nu} = \frac{1}{f}\int{ I(x,y)(\mu - \bar{\mu})(\nu - \bar{\nu})\dif{x} \dif{y}},
\end{equation}
where $\mu$ and $\nu$ refer to either $x$ or $y$, $\bar{\mu}$ and $\bar{\nu}$ indicate either the $x$ or $y$ centroids of the profile, and $f$ is the total flux of the profile.
In practice, when applying \ref{eqn:M0} - \ref{eqn:M2} to real imaging data, a weight function $w(x,y)$ is either explicitly (for moment-measuring algorithms) or implicitly (for model-fitting algorithms) included as a factor in each of the integrands to prevent noise in pixels at large radii from causing the integrals to diverge.
The second central moments of interest for weak lensing are those of galaxies before convolution with the PSF of the atmosphere, optics, and sensors.
The observed moments (after PSF convolution) are related to the intrinsic moments by $I_{\mu \nu}^\mathrm{obs} = I_{\mu \nu}^\mathrm{gal} + I_{\mu \nu}^\mathrm{PSF}$, which holds exactly only for unweighted second moments, but is still a useful approximation (correct to within a factor of a few) when using practical weight functions or model-fitting approaches.

Important combinations of second moments for weak lensing include the second-moment squared radius $r^2$, and the complex ellipticity $\boldsymbol{\chi} = \chi_1 + \mathrm{i}\chi_2$:
\begin{equation}
  \label{eqn:r2}
  r^2 = I_{xx} + I_{yy},
\end{equation}
\begin{equation}
  \label{eqn:chi1}
  \chi_1 = \frac{I_{xx} - I_{yy}}{I_{xx} + I_{yy}},
\end{equation}
\begin{equation}
  \label{eqn:chi2}
  \chi_2 = \frac{2 I_{xy}}{I_{xx} + I_{yy}}.
\end{equation}
Under the assumption that galaxy ellipticities are intrinsically isotropically distributed, the cosmological signal of interest, the reduced shear $\boldsymbol{g}$, is related to the ensemble average ellipticity by $\left\langle \boldsymbol{\chi^{(a)}} \right\rangle \approx 2 \boldsymbol{g}$.

Chromatic PSF effects are a subset of a larger class of effects in which one misestimates the size or shape of the PSF to be deconvolved from galaxy images.
Mathematically, we can describe this PSF misestimate via the error in the second moments: $\Delta I_{\mu \nu}^\mathrm{PSF} \equiv I_{\mu \nu}^\mathrm{PSF,g} - I_{\mu \nu}^\mathrm{PSF,*}$,
where $I_{\mu \nu}^\mathrm{PSF,g}$ are the true PSF moments appropriate for deconvolving a particular galaxy and $I_{\mu \nu}^\mathrm{PSF,*}$ are the PSF moments estimated from one or more stars.

To characterize the impact of PSF misestimation, we follow the literature and parameterize the bias induced in the reduced shear via a multiplicative and additive term:
\begin{equation}
  \label{eqn:shearbias}
  \hat{g}_i=g_i (1+m_i)+c_i,\quad i=1,2,
\end{equation}
where $\hat{g}$ is the estimator for the true reduced shear $g$, and we have assumed that $\hat{g_1}$ ($\hat{g_2}$) is independent of $g_2$ ($g_1$).
The results of propagating PSF misestimates into the multiplicative and additive shear biases are \cite{Meyers+Burchat15}
\begin{equation}
  \label{eqn:m1m2}
  m_1=m_2=\frac{-\left(\Delta I_{xx}^\mathrm{PSF}+\Delta I_{yy}^\mathrm{PSF}\right)}{r^2_\mathrm{gal}},
\end{equation}
\begin{equation}
  \label{eqn:c1}
  c_1=\frac{\Delta I_{xx}^\mathrm{PSF}-\Delta I_{yy}^\mathrm{PSF}}{2 r^2_\mathrm{gal}},
\end{equation}
\begin{equation}
  \label{eqn:c2}
  c_2=\frac{\Delta I_{xy}^\mathrm{PSF}}{r^2_\mathrm{gal}}.
\end{equation}
The typical source galaxy size for LSST is about $r^2_\mathrm{gal} \sim (0.3~\mathrm{arcsec})^2$ \cite{Meyers+Burchat15}.

Requirements on how well one needs to know $m$ and $c$ have been derived by several authors for surveys similar to and including LSST \cite{Huterer++06, Amara+Refregier08, Massey++13, Cropper++13}.
Interpreting these requirements can be tricky, however, and we refer the reader to Ref. \cite{Meyers+Burchat15} for more detailed discussion.
Following Ref. \cite{Meyers+Burchat15}, we set $\langle m \rangle _\mathrm{req} = 0.003$ (that is, the requirement on our knowledge of the mean of $m$ in each tomographic redshift bin), as the point at which the systematic uncertainties of LSST weak lensing become equal to the statistical uncertainties.
Similarly, we set $\Var{(c)}_\mathrm{req}=1.7\times10^{-7}$ as a conservative upper bound on the point at which uncorrected additive systematic uncertainties equal statistical uncertainties for LSST weak lensing.

\section{Toy models}\label{sec:toys}

In this section we construct a toy model for chromatic CCD PSF effects.
To isolate chromatic CCD effects from other types of chromatic effects, in our model we will assert that the PSF contribution coming from the atmosphere and telescope optics is achromatic.
For LSST, the atmospheric+optical PSF is dominated by the atmosphere, which we will model as a Moffat profile with FWHM of 0.65-arcsec and $\beta$ parameter of 5, which corresponds to $I_{xx} = I_{yy} = 0.12~\mathrm{arcsec}^2$.
We model the typical contribution of the CCD to the PSF as a Gaussian with FWHM of 0.19~arcsec, which is appropriate for charge diffusion in the LSST CCDs \cite{O'Connor++06, Nomerotski14, Fisher-Levine14}, and corresponds to $I_{xx} = I_{yy} = 0.0065~\mathrm{arcsec}^2$.

Mathematically, our model is
\begin{equation}
  I^\mathrm{PSF}_{\mu\nu}(\lambda) = I^\mathrm{PSF, telescope+optics}_{\mu\nu} +   I^\mathrm{PSF,CCD}_{\mu\nu}(\lambda).
\end{equation}
For a given SED, we can compute an effective PSF as
\begin{equation}
  \label{eq:PSFeff}
  I^\mathrm{PSF,eff}_{\mu\nu} \propto \int I^\mathrm{PSF}_{\mu\nu}(\lambda) R(\lambda) S(\lambda) \lambda \dif{\lambda},
\end{equation}
where $R(\lambda)$ is the total system throughput, $S(\lambda)$ is the SED, and the normalization is set such that the integral of the effective PSF over all space is 1.
Eq. \ref{eq:PSFeff} can then be used to compute the PSF misestimate and resulting multiplicative or additive shear bias for a given pair of stellar and galactic SEDs.

\subsection{Wavelength-dependent CCD PSF size}\label{subsec:chromatic_size}

To make the CCD PSF chromatic, we will slowly vary the second moments about their fiducial values as a function of wavelength.
Our first toy model varies the size of the CCD PSF as a function of wavelength:
\begin{equation}
  I_{xx}^\mathrm{PSF,CCD}(\lambda) = I_{xx}^\mathrm{PSF,CCD}(\lambda_0) + \beta_+ (\lambda-\lambda_0),
\end{equation}
\begin{equation}
  I_{yy}^\mathrm{PSF,CCD}(\lambda) = I_{yy}^\mathrm{PSF,CCD}(\lambda_0) + \beta_+ (\lambda-\lambda_0).
\end{equation}
We choose $\lambda_0 = 700$ nm to fall in between the LSST $r$- and $i$-bands.
To quantify the sensitivity of LSST shape measurement to a wavelength-dependent CCD PSF size, we choose $\beta_+ = 10^{-5}~\mathrm{arcsec}^2/$nm, which produces a CCD PSF with FWHM about 10\% smaller at the blue edge of the $i$-band filter than at the red edge.
This roughly corresponds to $\dif{(\mathrm{FWHM_{CCD}})}/\dif{\lambda} \sim 0.14~\mathrm{mas}/$nm, which, as we will see in Section~4, is about the right size for the LSST $i$-band.

Since this model changes only the sum $I_{xx} + I_{yy}$, but leaves the difference $I_{xx} - I_{yy}$ and $I_{xy}$ fixed, it generates only a multiplicative bias.
Figure 1 shows the magnitude of this multiplicative bias relative to a G5v star for a variety of stellar and galactic SEDs.
The shaded region in the plot indicates the LSST requirement for how well we need to know the average multiplicative bias in each tomographic redshift bin so that systematic uncertainties remain at or below statistical uncertainties.
Since the different SEDs roughly span this requirement range, we conclude that $\beta_+ \sim 10^{-5}~\mathrm{arcsec}^2/$nm or $\dif{(\mathrm{FWHM_{CCD}})}/\dif{\lambda} \sim 0.14~\mathrm{mas}/$nm describes the point at which the uncertainty from this systematic effect rivals the statistical uncertainty of LSST weak lensing.

\begin{figure}
  \label{fig:betaplus}
  \includegraphics[width=1.0\textwidth]{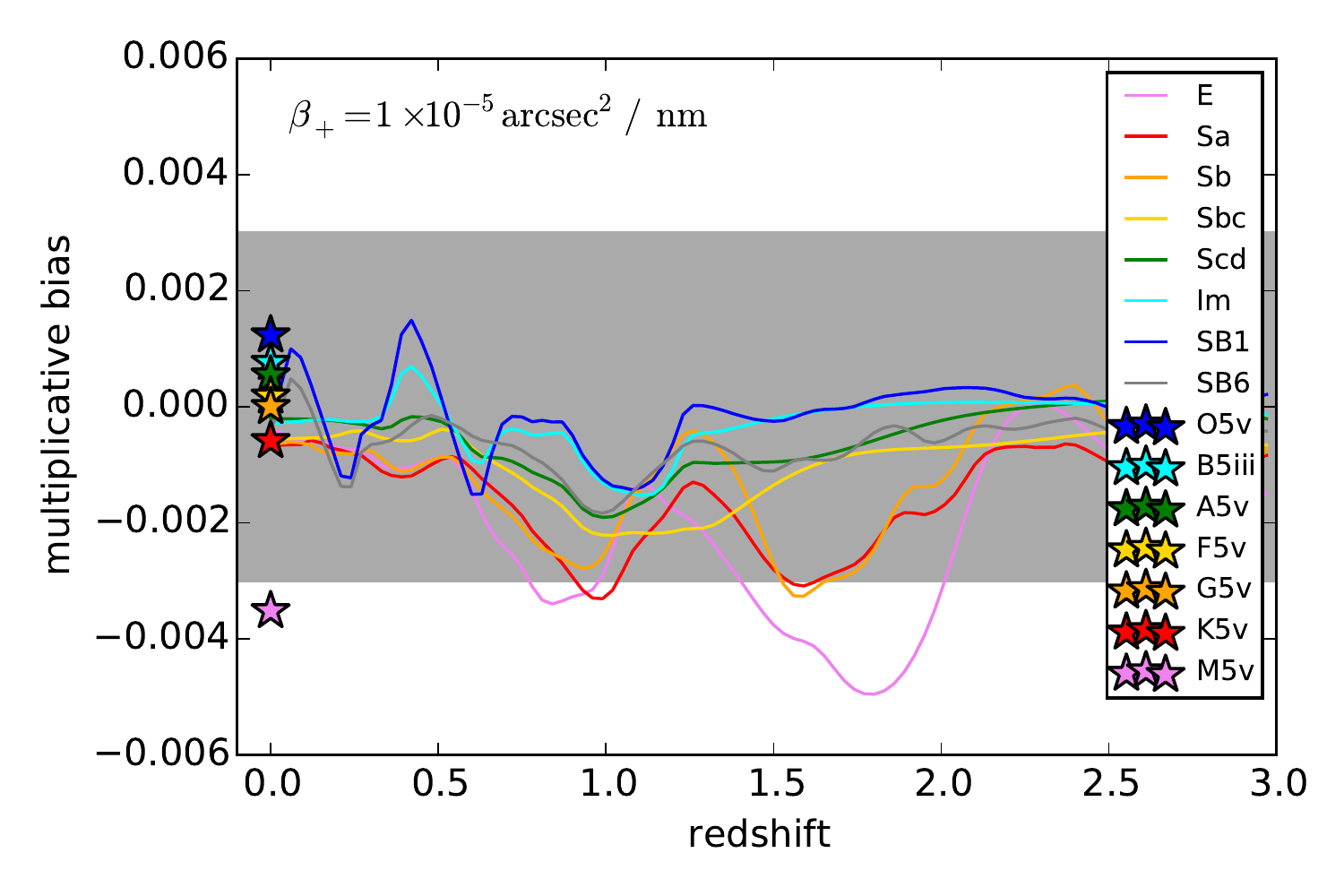}
  \caption{
    Multiplicative shear bias arising from a toy model with linear wavelength dependence $\dif{I^\mathrm{PSF,CCD}_{xx}}/\dif{\lambda} = \dif{I^\mathrm{PSF,CCD}_{yy}}/\dif{\lambda} = \beta_+$, calculated for LSST $i$-band.
    Star symbols at redshift 0 represent stellar SEDs from \cite{Pickles98}.
    Curves represent galactic SEDs from \cite{Coleman++80} and \cite{Kinney++96}.
    The gray band indicates how well one needs to know the multiplicative bias in each LSST tomographic redshift bin so that systematic uncertainties are smaller than statistical uncertainties.
    The biases and requirement are similar in the $r$ band, which is the other band planned for shape measurement with LSST.
  }
\end{figure}

\subsection{Wavelength-dependent CCD PSF ellipticity}\label{subsec:chromatic_ellipticity}

We also construct a somewhat orthogonal toy model in which the ellipticity, but not the size, of the CCD PSF is chromatic:
\begin{equation}
  I_{xx}^\mathrm{PSF,CCD}(\lambda) = I_{xx}^\mathrm{PSF,CCD}(\lambda_0) + \beta_- (\lambda-\lambda_0),
\end{equation}
\begin{equation}
  I_{yy}^\mathrm{PSF,CCD}(\lambda) = I_{yy}^\mathrm{PSF,CCD}(\lambda_0) - \beta_- (\lambda-\lambda_0).
\end{equation}
Using the same $\lambda_0=700$~nm as before and setting $\beta_- = 10^{-5}~\mathrm{arcsec}^2/$nm results in a CCD PSF that changes in ellipticity from about $-0.1$ at the blue edge of the $i$-band to about $+0.1$ at the red edge of the $i$-band.
This is roughly equivalent to $\dif{\chi_{1,\mathrm{CCD}}}/\dif{\lambda} \sim 0.0014~\mathrm{nm}^{-1}$, which, as we will see in Section~4, is a factor of a few smaller than observed in LSST $i$-band simulations.

This model induces an additive bias in the $\chi_1$ ellipticity component, but not in the $\chi_2$ ellipticity component.
It does not induce a multiplicative bias in either ellipticity component.
Figure 2 shows the magnitude of this additive bias relative to a G5v star for a variety of stellar and galactic SEDs.
The shaded region in the plot indicates the LSST sufficient-but-not-necessary requirement for how well we need to constrain the standard deviation of the additive bias in each tomographic redshift bin so that systematic uncertainties remain at or below statistical uncertainties.
In this case, we see that the requirement bar is a factor of a few smaller than the range spanned by the various SEDs.
We therefore conclude that $\beta_- \sim 3\times10^{-6}~\mathrm{arcsec}^2/$nm or equivalently $\dif{\chi_{1,\mathrm{CCD}}}/\dif{\lambda} \sim~5~\times~10^{-4}~\mathrm{nm}^{-1}$, describes an upper limit to the point at which the uncertainty from this systematic effect could rival the statistical uncertainty of LSST weak lensing.

\begin{figure}
  \label{fig:betaminus}
  \includegraphics[width=1.0\textwidth]{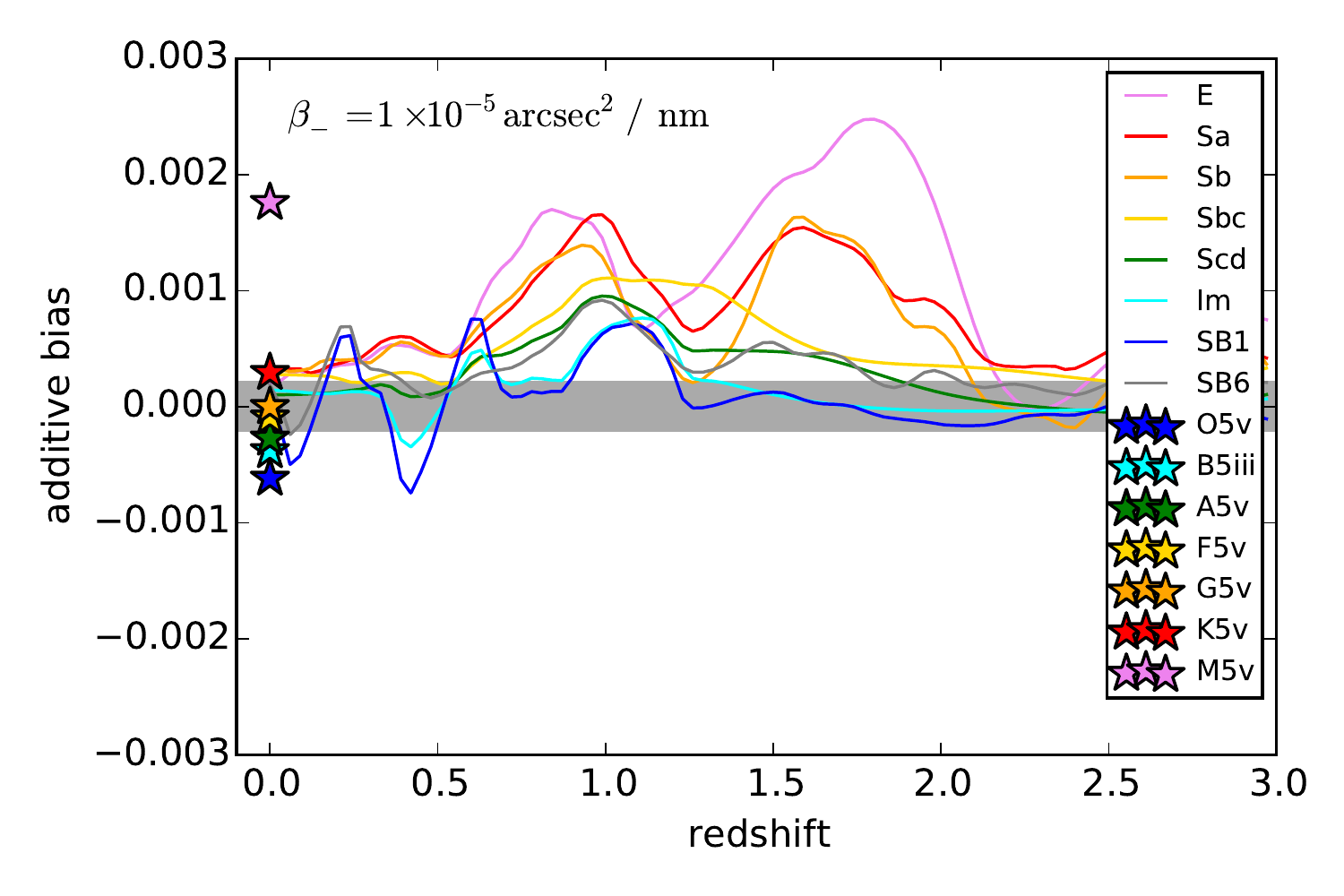}
  \caption{
    Additive shear bias arising from toy model with linear wavelength dependence $\dif{I^\mathrm{PSF,CCD}_{xx}}/\dif{\lambda} = -\dif{I^\mathrm{PSF,CCD}_{yy}}/\dif{\lambda} = \beta_-$, calculated for LSST $i$-band.
    Star symbols at redshift 0 represent stellar SEDs from \cite{Pickles98}.
    Curves represent galactic SEDs from \cite{Coleman++80} and \cite{Kinney++96}.
    The gray band indicates the sufficient requirement for LSST -- i.e. the point at which systematic uncertainties will definitely not exceed statistical uncertainties.
    The biases and requirement are similar in the $r$ band, which is the only other band planned for shape measurement with LSST.
    }
\end{figure}

\section{PhoSim estimates of LSST CCD chromaticity}\label{sec:phosim}

To connect our toy models to the actual LSST survey, we use the LSST photon simulator (\textsc{PhoSim}) \cite{Peterson++15}, which comprises one part of the LSST simulations package.
\textsc{PhoSim} works by drawing photons from sources in a provided catalog, and then manipulating them one-by-one as they traverse the atmosphere, telescope and camera optics, and convert into electron-hole pairs inside the LSST sensors with a wavelength-dependent absorption length \cite{Rajkanan++79}.
These photo-electrons are then further followed as they diffuse towards the charge collection areas of each simulated pixel and are eventually read out by the simulated sensor electronics.
In addition to its use as a high-fidelity simulator of LSST before the actual telescope is available, \textsc{PhoSim} is a valuable resource for probing individual physical effects, which can be switched on and off on demand in the software.

Here we use \textsc{PhoSim} to probe chromatic PSF effects in the LSST sensors.
We create a series of catalogs, each of which consists of a $8\times8$ grid of stars with spacing 0.01~degrees.
All of the stars in a given catalog have the same monochromatic SED -- i.e., the flux is non-zero at only a single wavelength.
This wavelength is then scanned from 325~nm to 1050~nm in steps of 25~nm to produce different catalogs.
We tuned the amplitude of each monochromatic SED such that the resulting stellar images each contained about 20,000 counts.
We replicate this pattern of stars at three different field angles: on axis, 0.725~degrees off axis (about half way from the center to the edge of the focal plane), and 1.45~degrees off axis (near the edge of the focal plane).
The off-axis positions are azimuthally positioned along the positive ``x''-direction.
The zenith angle of the telescope boresite is always set to an altitude of 89~degrees in the simulation, which is close enough to the zenith that differential chromatic refraction is negligible across the entire field of view.

To isolate effects arising in the sensors, we run \textsc{PhoSim} for each input catalog in two different modes: once with the atmosphere, optics, and sensors all turned on, and once with only the atmosphere and optics turned on.
To simplify our analysis, we also turn off a number of additional physics effects, including the sky background, saturation, blooming, tracking errors, optics perturbations, clouds, quantum-efficiency variations, optical surface imperfections (dust), diffraction, large angle scattering, lateral electric fields (e.g., tree rings and edge effects), charge sharing (i.e., the brighter-fatter effect), imperfect pixel boundaries, and shutter errors.

In the resulting images, we fit a Gaussian profile with six free parameters ($x_0$, $y_0$, $f$, $\chi_1$, $\chi_2$, FWHM) to each star.
(We also tried fitting a Moffat profile to the stars; the results are essentially the same.)
For each wavelength and mode, we store the mean and standard deviation of the parameters of the 64 stars in the $8\times8$ grid.
In Figure 3, we plot the quadrature difference of the mean FWHM with the sensors turned on and with the sensors turned off as a function of wavelength and for the three different field angles.
By plotting the quadrature difference, we are isolating the contribution to the PSF from the CCDs at each wavelength.
We see that CCD PSF FWHM is constant at bluer wavelengths before rising around 800~nm, roughly independent of field angle.
In Table \ref{table:phosim}, we list the measured values of $\dif{(\mathrm{FWHM_{CCD}})}/\dif{\lambda}$ for each filter and field angle.
We see that in $i$-band, the measured slope is only about a factor of two to three smaller than the requirement we established in Section \ref{subsec:chromatic_size}.
The slope is smaller in $r$-band, though the uncertainties on the measurements are significant here.

\begin{figure}
  \label{fig:fwhm}
  \includegraphics[width=1.0\textwidth]{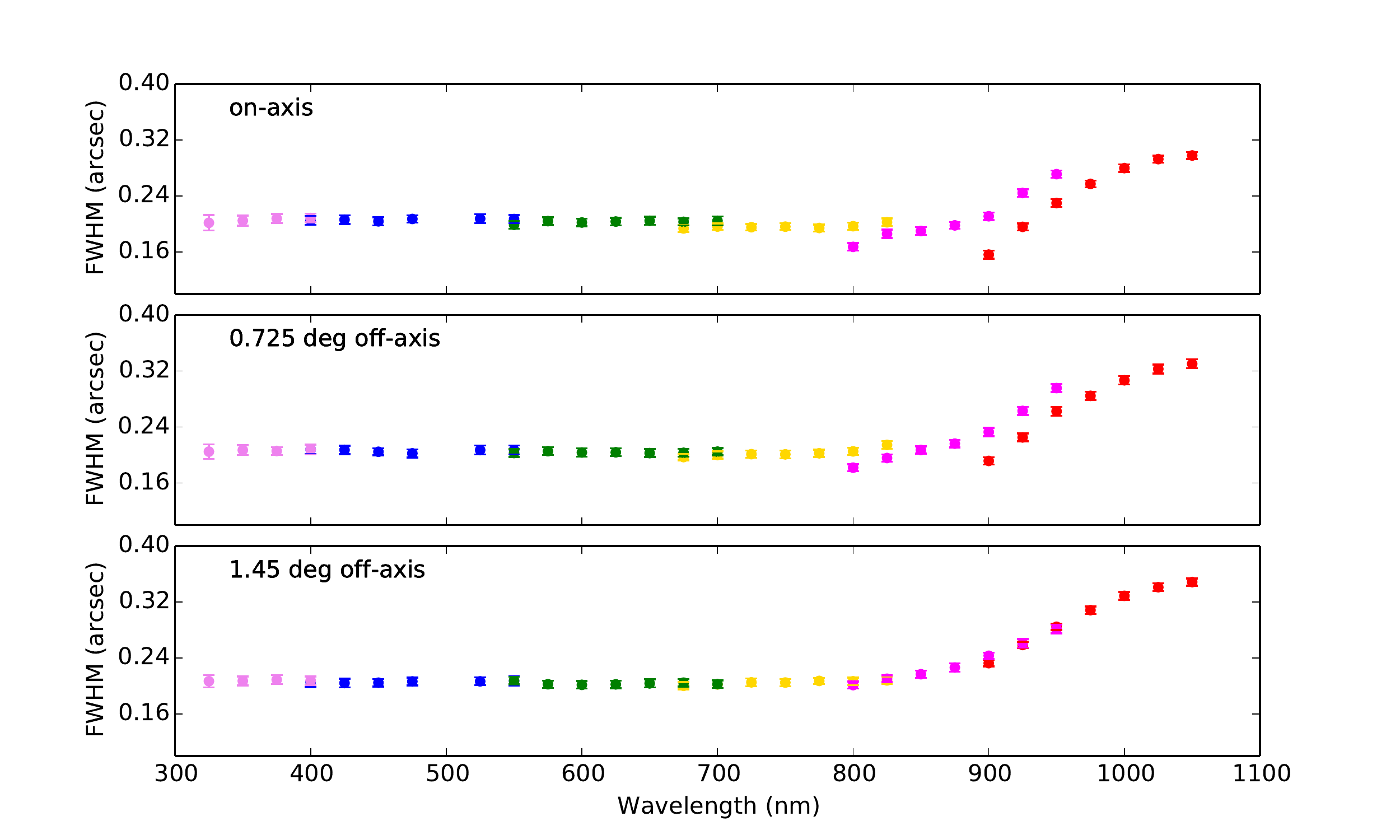}
  \caption{
    The FWHM of the LSST CCD PSF estimated using \textsc{PhoSim} as a function of wavelength and for three different field angles.
    Different colors indicate different simulated LSST filters (ugrizy).
  }
\end{figure}

To investigate the CCD PSF ellipticity, we convert our measurements of $\chi_1$, $\chi_2$, and $\mathrm{FWHM}$ into second central moments $I_{\mu\nu}$, subtract the second central moments with and without the sensor turned on, and convert the residuals back into $\chi_1$ and $\chi_2$.
Figure 4 shows the $\chi_1$ ellipticity component against wavelength for several field angles (i.e., for several angular distances from the center of the focal plane), and for an azimuthal direction of displacement along the ``+x'' direction of the focal plane.
We see that the CCD PSF ellipticity is nearly constant on-axis, but becomes progressively larger with wavelength off-axis.
At this azimuthal angle, the $\chi_2$ residual component (not plotted) is uniformly consistent with zero for any field angle.
Repeating the above experiment for azimuthal angles along the ``x=y'' direction (``+y'' direction) we find that the $\chi_1$ ($\chi_2$) ellipticity is zero and $\chi_2$ ($\chi_1$) ellipticity is positive (negative) off axis, which strongly implies that the effect is radially symmetric in the focal plane.
In Table \ref{table:phosim}, we list the measured values of $\dif{\mathrm{\chi_{1,CCD}}}/\dif{\lambda}$ for each filter and field angle for the case when the azimuthal displacement is along the ``+x'' direction.
While the slope on-axis is consistent with zero, off axis the slope is a factor of one to three larger than our requirement of $\sim 0.5 \mu$m$^{-1}$ from Section \ref{subsec:chromatic_ellipticity}.

\begin{figure}
  \label{fig:chi1}
  \includegraphics[width=1.0\textwidth]{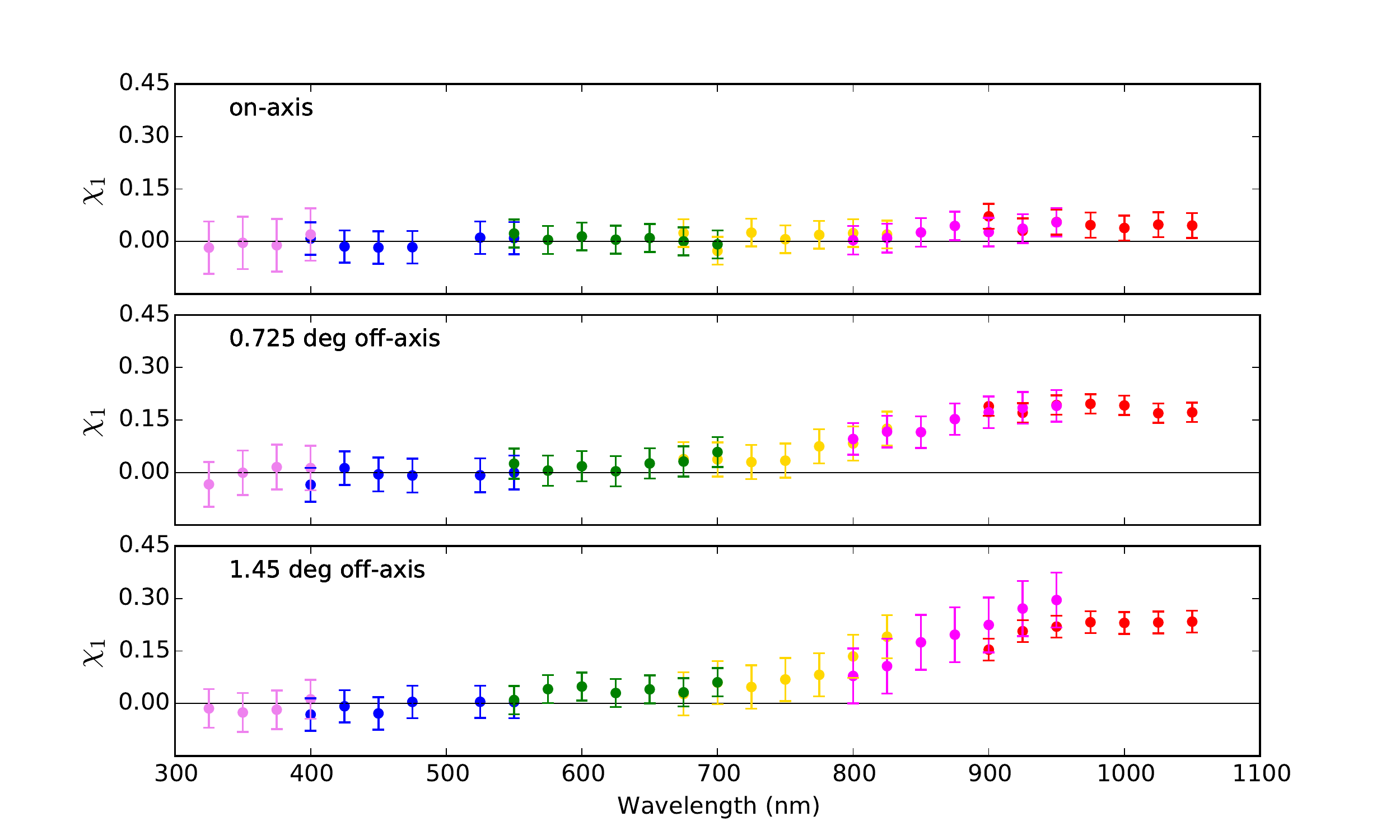}
  \caption{
    Ellipticity $\chi_1$ of the LSST CCD PSF estimated using \textsc{PhoSim} as a function of wavelength and for three different field angles.
    Different colors indicate different simulated LSST filters (ugrizy).
  }
\end{figure}

\begin{table}
    \centering
    \caption{Estimated LSST CCD PSF chromaticity}
    \smallskip
    \begin{tabular}{|lcccccc|}
        \hline
        Filter      & \multicolumn{3}{c}{$\dif{(\mathrm{FWHM_{CCD}})}/\dif{\lambda}~~~(\mathrm{mas}/\mathrm{nm})$} & \multicolumn{3}{c}{$\dif{\mathrm{\chi_{1,CCD}}}/\dif{\lambda}~~~(\mu\mathrm{m}^{-1})$} \\
                    & 0.0$^\circ$       & 0.725$^\circ$    & 1.45$^\circ$       & 0.0$^\circ$      &  0.725$^\circ$   &  1.45$^\circ$ \\
        \hline
        $u$         & 0.08 $\pm$ 0.03  & 0.03 $\pm$ 0.03  &  0.02 $\pm$ 0.03  &  0.4 $\pm$ 0.3  &  0.6 $\pm$ 0.3  & 0.4 $\pm$ 0.4 \\
        $g$         & 0.02 $\pm$ 0.02  & 0.00 $\pm$ 0.04  &  0.01 $\pm$ 0.02  &  0.1 $\pm$ 0.2  &  0.1 $\pm$ 0.3  & 0.2 $\pm$ 0.2 \\
        $r$         & 0.03 $\pm$ 0.03  & 0.00 $\pm$ 0.02  & -0.02 $\pm$ 0.04  & -0.2 $\pm$ 0.1  &  0.2 $\pm$ 0.3  & 0.2 $\pm$ 0.2 \\
        $i$         & 0.04 $\pm$ 0.04  & 0.09 $\pm$ 0.05  &  0.05 $\pm$ 0.02  &  0.1 $\pm$ 0.3  &  0.6 $\pm$ 0.3  & 1.0 $\pm$ 0.4 \\
        $z$         & 0.65 $\pm$ 0.20  & 0.71 $\pm$ 0.18  &  0.51 $\pm$ 0.11  &  0.3 $\pm$ 0.2  &  0.7 $\pm$ 0.2  & 1.5 $\pm$ 0.2 \\
        $y$         & 0.94 $\pm$ 0.24  & 0.96 $\pm$ 0.19  &  0.79 $\pm$ 0.14  & -0.1 $\pm$ 0.2  & -0.1 $\pm$ 0.2  & 0.4 $\pm$ 0.3 \\
        \hline
    \end{tabular}
    \label{table:phosim}
\end{table}

\section{Discussion}\label{sec:discussion}

We expect the response of CCDs to be chromatic due to the wavelength-dependent absorption length of silicon.
Redder photons penetrate farther into the silicon on average before converting into electron-hole pairs.
Naively, one might then conclude that photo-electrons generated by redder photons would on average have less opportunity to laterally diffuse and would ultimately produce a smaller PSF.
However, Figure 3 suggests that the opposite is true.
The naive argument above neglects the fact that the LSST beam is extremely fast -- f/1.2.
Photons arrive at the CCD with a mean angle of incidence around 20 degrees, with a range of $\approx \pm 5$~degrees due to the annular primary mirror, which leads to a mean angle of refraction around 5~degrees inside the silicon.
As shown in Figure 5, blue photons convert into electron-hole pairs almost immediately, and the photo-electrons diffuse across the entire depth of the CCD.
Redder photons convert further inside the CCD on average, which, combined with the lateral component of their momentum, yields a larger PSF.
Note that the position of best focus is inside the bulk of the CCD for redder filters as a result \cite{O'Connor++06}.

\begin{figure}
  \label{fig:ccd}
  \includegraphics[width=1.0\textwidth]{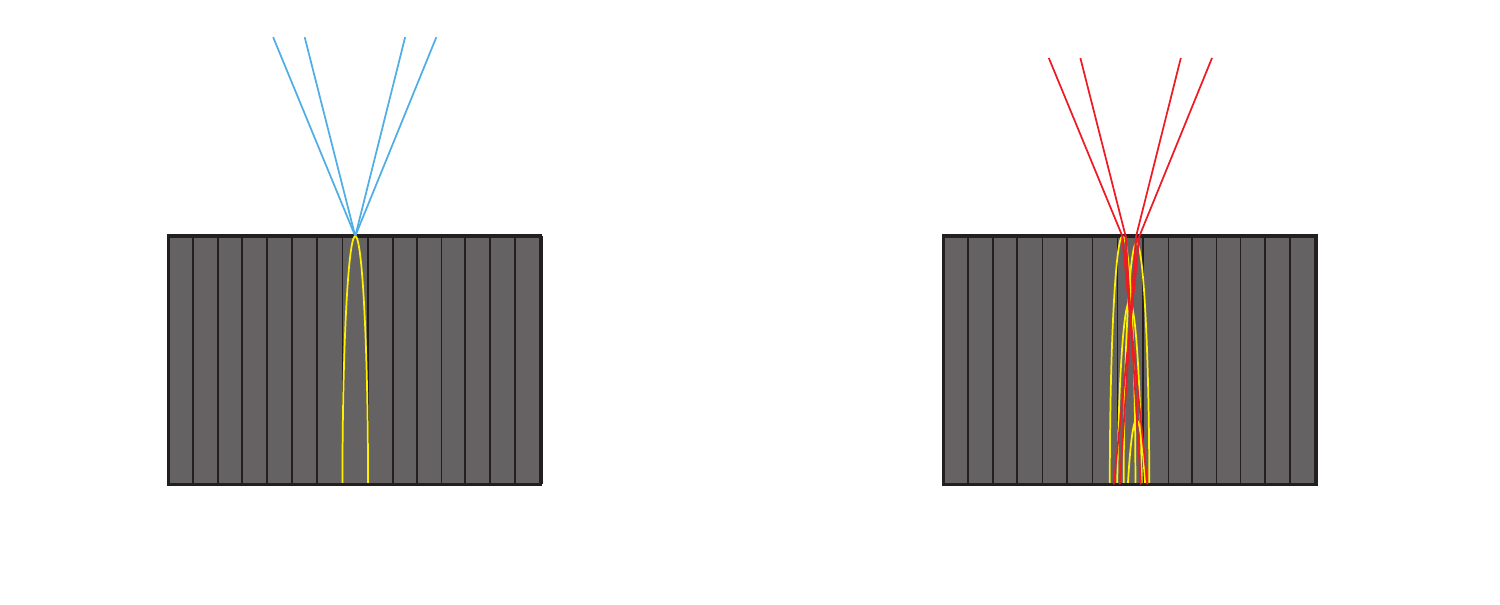}
  \caption{
    Cartoon figure of LSST beam and CCD geometry.
    Fifteen 10-micron wide and 100-micron deep pixels are shown in each cartoon.
    The f/1.2 beam is shown approaching the CCD from above for bluer filters (left) and redder filters (right).
    Blue photons convert into electron-hole pairs almost immediately, and the photo-electrons diffuse the entire depth of the CCD (shown in yellow).
    Red photons may penetrate deeper into the CCD bulk before converting.
    Combined with the diverging beam, which is still f/4.5 inside the silicon, the longer absorption length of redder photons leads to a larger PSF.
    Note that the position of best focus is a competition between beam divergence and charge diffusion.
    For bluer filters, the best focus is at the CCD surface, while for redder filters it is inside the bulk of the CCD \cite{O'Connor++06}.
    }
\end{figure}

Similarly, even the chief ray of the LSST beam (though blocked by the secondary mirror and camera) is inclined to the CCD normal at an off-axis field position.
In other words, at an off-axis field position, the unblocked rays do not average together to normal incidence.
This is likely the source of the wavelength-dependent ellipticity at off-axis field positions in Figure 4.

Although the CCD contribution to the LSST PSF is subdominant to the atmospheric contribution by a factor of a few in FWHM, both effects we have investigated here appear to be large enough to introduce significant biases in cosmic shear results if left uncorrected.
Higher-order CCD effects, such as tree-rings, charge sharing, or edge effects may also be chromatic and must be investigated, especially in combination with the effect investigated here and with atmospheric and optical chromatic effects, since there may be non-trivial interactions among the different effects.

If the SED of each object and the wavelength dependence of the PSF are known, then one can compute potential chromatic PSF misestimations and remove them from the analysis, galaxy-by-galaxy and star-by-star.
Fortunately, the six-band photometry that LSST will obtain of each source will significantly constrain the SED across the $r$- and $i$-band filters where shape measurements are most effective.
In Ref. \cite{Meyers+Burchat15}, we have shown that a significant reduction in bias can be achieved by training a machine-learning algorithm to predict chromatic biases from photometry.
Such a correction requires an accurate and precise knowledge of the wavelength dependence of the PSF, however, which motivates continued study of chromatic effects both in simulation and in data where possible.

\acknowledgments
This work was supported by National Science Foundation grant PHY-1404070.

\end{document}